\documentclass[letterpaper, 10 pt, conference]{ieeeconf}
\IEEEoverridecommandlockouts                              
\usepackage{subcaption}
\usepackage[utf8]{inputenc}
\PassOptionsToPackage{dvipsnames, table}{xcolor}
\usepackage{color}
\usepackage{array}
\usepackage{verbatim}
\usepackage{float}
\usepackage{multirow}
\usepackage{amsmath}
\usepackage{amssymb}
\usepackage{tabularx}
\usepackage{slashed}
\usepackage{svg}
\usepackage{ifsym}
\usepackage{wasysym}
\usepackage{tikz}
\usepackage{pgfplots}
\pgfplotsset{compat=1.17}
\usepackage{caption}
\usepackage{wrapfig}
\usepackage{diagbox}
\usepackage{booktabs}
\usepackage{cite}
\usepackage{url}
\usepackage{color}
\usepackage{humberto_commands}
\usepackage{humberto_variable_names}
\usepackage[dvipsnames]{xcolor}

\title{\LARGE \bf Information-Aware Guidance for Magnetic Anomaly based Navigation}
\author{J. Humberto Ramos$^{1}$, Jaejeong Shin$^{1}$, Kyle Volle$^{1}$, Paul Buzaud$^{1}$, Kevin Brink$^{2}$, Prashant Ganesh$^{1}$
\thanks{*This work was partly supported by the Air Force Research Lab through contract FA8651-20-F-1025 }
\thanks{$^{1}$J. Humberto Ramos, Jaejeong Shin, Kyle Volle, Paul Buzaud and Prashant Ganesh are with the Department of Mechanical and Aerospace Engineering at the University of Florida.
        {\tt\small {jramoszuniga,jshin2,kvolle,paul.buzaud,}
        \tt\small {prashant.ganesh}@ufl.edu}}%
\thanks{$^{2}$Kevin Brink is with the Air Force Research Lab, Munitions Directorate at Eglin Air Force Base. {\tt\small kevin.brink@us.af.mil}}%
}
\begin{document}

\maketitle
\thispagestyle{empty}
\pagestyle{empty}
\begin{abstract}
    
    In the absence of an absolute positioning system, such as GPS, autonomous vehicles are subject to accumulation of positional error which can interfere with reliable performance. Improved navigational accuracy without GPS enables vehicles to achieve a higher degree of autonomy and reliability, both in terms of decision making and safety. This paper details the use of two navigation systems for autonomous agents using magnetic field anomalies to localize themselves within a map; both techniques use the information content in the environment in distinct ways and are aimed at reducing the localization uncertainty. The first method is based on a nonlinear observability metric of the vehicle model, while the second is an information theory based technique which minimizes the expected entropy of the system. These conditions are used to design guidance laws that minimize the localization uncertainty and are verified both in simulation and hardware experiments are presented for the observability approach.

\end{abstract}

\section{Introduction}
The ability of a robot to localize itself with respect to its environment is an essential component of any autonomous system. While the global navigation satellite system (GNSS) is used in many systems to provide sub-meter level accuracy position measurements, in some cases GNSS can be degraded or unreliable due to interference from the environment, or may be unavailable. Robot localization in GPS-denied environments has received significant attention in the recent years, particularly delving into solutions that blend information from alternative sensors to determine a navigation solution. Popular solutions for GPS-denied navigation include visual-aided inertial navigation systems which use a measurement unit (IMU) paired with a vision sensor (such as an RGB-D), \cite{koch2020relative, mur2015orb} a stereo camera \cite{7995765}, or a monocular camera \cite{usenko2016direct, joshi2019experimental}. It is also common to find techniques that fuse IMU data with lidar or event sensors information such as those presented in \cite{droeschel2018efficient, khan2021comparative}.
Although these methods have been shown effective to navigate in GPS-denied environments, they tend to struggle with visually uniform scenes, such as over water, where there are very few visual distinct features, and reflective sensors may fail. Moreover, camera-based systems are especially reliant on weather and lighting conditions which may complicate its use in outdoor missions, whereas in sparse environments like a desert or the sea, lidar-based systems may be limited by the maximum range they can sense.

To overcome basedsome of these challenges, emergent alternative systems for GPS-denied navigation based on magnetic anomalies have been explored in the last decades \cite{solin2016terrain, pritt2014indoor} as a viable solution; it is globally available and is independent of weather and lighting conditions. A common approach for localizing autonomous agents is to match on-board magnetometer measurements to prior magnetic maps using Bayesian estimation techniques such as a particle filter \cite{canciani2016absolute}. 

Along these lines, in this work we implement two approaches for localization-aware guidance using a scalar magnetometer and an a priori magnetic field map. Each approach pairs a metric for the vehicle's localization uncertainty with a local optimization algorithm for path planning to produce a navigation solution. The navigation solutions take into account the magnetic field information content of the environment in order to plan paths that maximize localization accuracy. The first approach uses a classical observability metric from control theory to maximize the gain of information along the path taken, while the path itself is generated via a receding-horizon dynamic programming approach. This approach allows for finding the locally optimal path with a search horizon that can be tuned to the computation capacity of the platform. The second approach uses the expected entropy of the system as the metric that, along with the distance to goal, is minimized at every time step. Due to the heavy computational burden that comes with this second method and how the computations scale, the planner looks ahead only a single time step. This trades consideration of future states for more tightly bound uncertainty. Maintaining bounded pose uncertainty is critical for long-term autonomy of agents as it impacts both safety and high-level planning decisions. The path planning problem is treated here as a local optimization problem to consider the presence of dynamic obstacles, actuator miscalibration, or sensor errors that can lead to trajectory changes or drift. 

Path planning using the information content of the environment has been widely studied in the community and is commonly referred to as coastal navigation \cite{roy1999coastal}. In \cite{yu2011observability} an observability-based path planning algorithm is used to avoid obstacles and simultaneously drive a vehicle to its goal position. This technique is demonstrated for small air vehicles using time-to-collision estimates and measurements of bearing to objects in the environment. Reference \cite{alaeddini2016optimal} introduces an optimal control problem that optimizes on the observability of a nonlinear system. Similarly, \cite{roy1999coastal} provides a framework for a probabilistic navigation algorithm that uses the entropy to plan trajectories for a robot. \cite{stachniss2005information} proposes combining simultaneous localization and mapping (SLAM) and exploration by using a Rao-Blackwellized particle filter to provide posterior pose estimates to feed into an information-based exploration algorithm. The work from \cite{van2011lqg} uses a priori knowledge of sensors to design optimal paths for a robot exploring an unknown space. Other recent works on localization-informed path planning have explicitly considered the use of magnetometers for geomagnetic feedback. These fall into three main categories: the first, and least applicable, are studies that look at path planning techniques used by animals to account for local anomalies that interfere with North/South navigation \cite{li2017geomagnetic, zein2021simulation}. The second category is to integrate the Cramer-Rao Lower Bound (CRLB) over the path, treating it as part of the path cost that is being minimized. By minimizing the CRLB, the uncertainty is minimized \cite{quintas2019auv}. The final category of methods discretizes the domain of interest and calculates a score for each cell that reflects how suitable it is for matching two magnetometer measurements, then the path is optimized to avoid these cells as much as possible \cite{teng2018dynamic, liu2020application}. Our approaches differ from the previous categories in that our algorithms optimize the information gain directly and locally.
The main contribution of this article includes the application of two distinct approaches to the magnetic anomaly-based navigation problem. Both the observability- and entropy-based techniques are shown to work in a high-fidelity simulation using a scalar magnetic anomaly map generated from repeated surveys of our lab space. Additionally, the observability-based approach is implemented on a differential-drive robot to validate the technique in a real-world scenario, using a particle filter to incorporate the magnetometer readings into the robot's state estimation.


This paper is organized as follows, Section~\ref{sec:prelim} contains preliminaries on the map collection process and the particle filter based localization algorithm. Section III details the two information-based navigation algorithms. Section IV contains results from experiments run on both simulated and hardware environments; note that for both the experiments a real magnetic map is used. Section V summarizes our findings and discusses some ideas for future work.

\section{Preliminaries}\label{sec:prelim}

This section provides some preliminaries on the process for collecting the indoor magnetic map of our laboratory space. Additionally, we include a brief description of the Monte Carlo navigation algorithm (particle filter) used in this work to demonstrate the proposed guidance techniques. 



\subsection{Indoor Magnetic Map Collection} \label{sec:MapCollection}

The map of our indoor space is collected using a scalar magnetometer mounted on the top of a ground differential-drive robot. This same configuration will be used for the hardware tests in Section \ref{sec:results}, so the recorded map will differ from the magnetic field measured at run time by as little as possible.

\begin{figure}[!htb]
    \centering
    \includegraphics[width=0.41\textwidth]{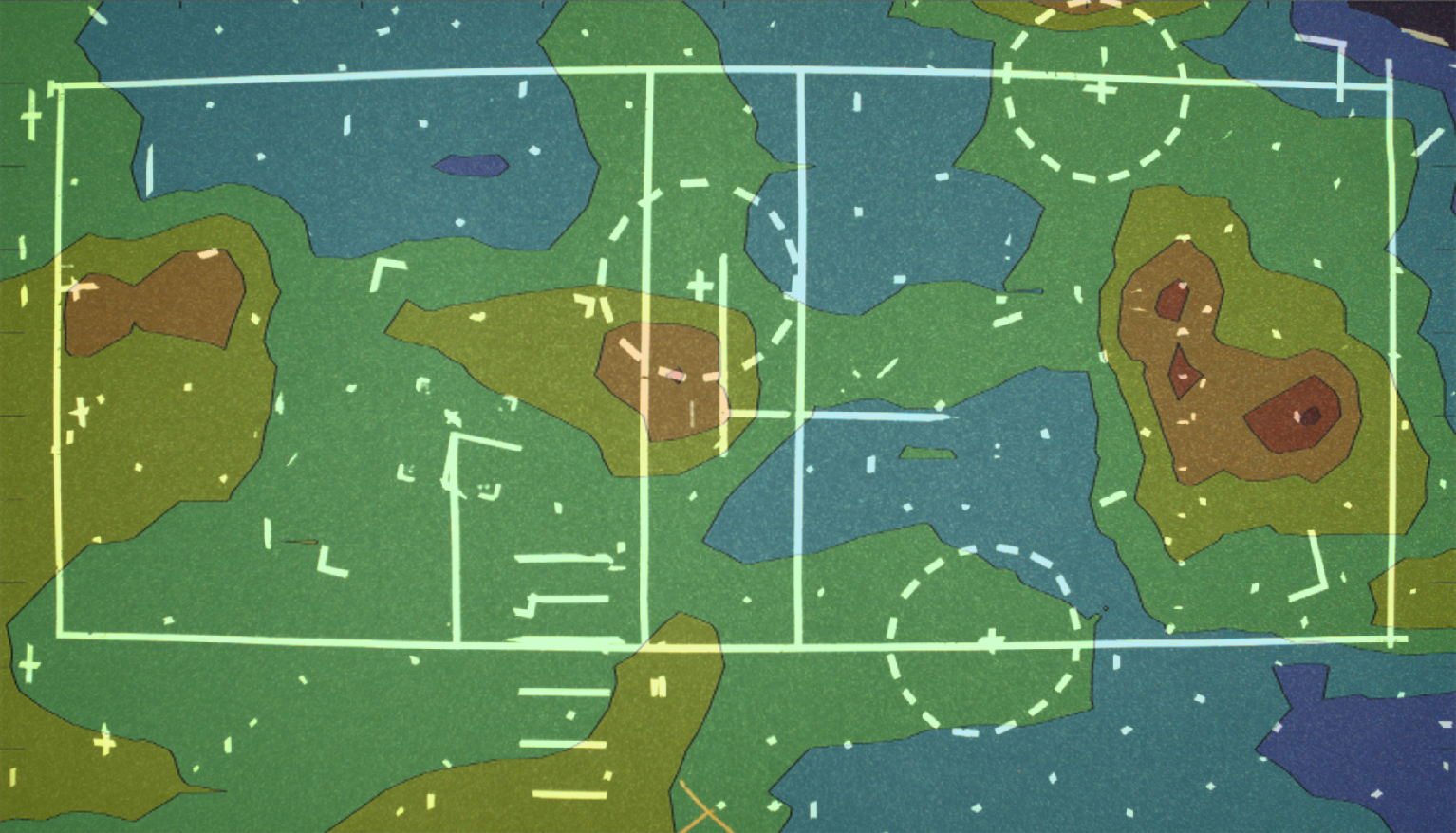}
    \caption{The north facing magnetic anomaly map overlaid over the laboratory space. The north facing map is composed by the magnetic measurements recorded with the ground robot facing the Earth magnetic north.}
    \label{fig:map_contour}
\end{figure}

For the data collection, a motion capture system is used to gather ground position data and to provide pose feedback for the ground robot control. The robot stops at every node of a grid of 25 cm x 25 cm to collect magnetic data. Stopping the robot at each location reduces the effect that actuators can have on the baseline map generation. 
\begin{figure}[!htb]
    \centering
    \includegraphics[width=0.3\textwidth]{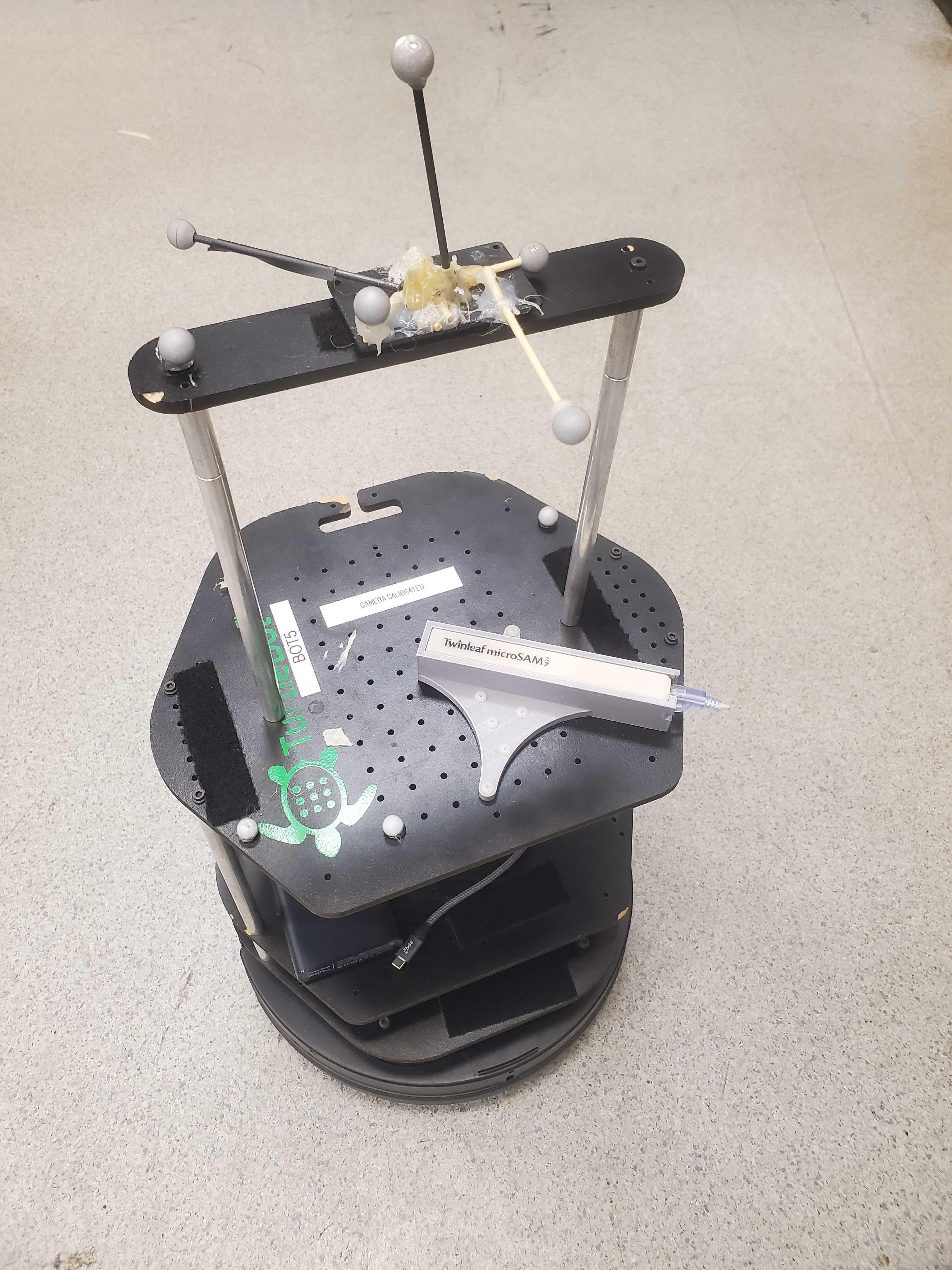}
    \caption{Ground robot used to collect the magnetic anomlay map using a TwinLeaf Microsam scalar magnetometer.}
    \label{fig:robotic_plaform}
\end{figure}
In addition to external disturbances, the heading of the robot also affects the magnetic field intensity measured at each location as the robot itself interacts with the total field. In this article, we use the magnetic north facing map as the baseline map and model the change in magnetic field, due to heading changes, as a sinusoidal function at a given location. Fig. \ref{fig:map_contour} shows the magnetic map contours of our laboratory space.  

Magnetic maps of the Earth's crustal field are widely available through various federal agencies and are regularly updated even though the crustal field remain approximately constant on a geological time scale. Mapping an indoor environment can be challenging as the magnetic field intensity in a given area is affected by electronics such as computers or even air conditioning systems. Fortunately, the effect of a disturbance to the magnetic field is inversely proportional to the third power of the distance from the disturbance. For our experiments the optically pumped magnetometer was placed at the top of the robot with the intent of minimizing disturbances from onboard components. The
consistency of the magnetic map was verified with additional data collections that were done twice, three-months apart, showing an average of 3$\%$ error over the whole mapped space. In order to foster collaborations, we have made this map publicly available at the University of Florida Autonomous Vehicles Laboratory website. 
\subsection{Particle Filter for Localization}\label{sec:particle_filter}
Localization of a vehicle using a standard particle filter is essentially comprised of two major steps. In the first step, a sample of the vehicle state space $\mathcal{X}^s$ is obtained at instant $k$ and propagated in time via the vehicle dynamics model. This sample has elements (particles) $\xk^i$ that may cover all possible ranges of the system state-space, or only a sub-set near an initial state guess, if available. In this article, the localization task using scalar magnetometer measurements is done for a differential robot where the discrete-time kinematic unicycle model, $\x_{k+1}=\stdvec{f}(\xk,\uk)$, for inertial pose, $\xk=[x_1,x_2,x_3]\trans=[x,y,\theta]\trans$, with forward velocity and angular rate inputs, $\uk=\trans[V,\omega]$, and integration interval, $\Delta t$, is used for simplicity:  
\begin{equation}
    \label{eq:dynamics}
    \begin{array}{c}
        x_{k+1}= x_k+ \mathrm{V} \cos \theta\Delta t \\
        y_{k+1}= y_k+\mathrm{V} \sin \theta\Delta t \\
        \theta_{k+1}=\theta_k+\omega\Delta t
    \end{array}
\end{equation}
In the second major step of the particle filter, an importance weight $\w_i$ is computed and associated to each state $\xk^i$ in the sample. The weight $\w_i$ takes on a value commensurate to the likelihood of the associated state $\xk^i$ being the true state $\x_t$. The value of $\w_i$ depends on the residual $r$ between the predicted sensor measurement $h_i(\xk^i)$ and the sensor reading $\tilde{y}$ at the current time $k$. This is, 
\begin{equation}\label{eq:residual}
    r^i_k = \tilde{y}_k - h_i(\xk^i)
\end{equation}

Particles that produce small residuals $r$, receive higher weights as they are more likely to be the actual state of the vehicle. On the other hand, particles with large residuals are less likely to represent the actual state of the system, and thus, low weights are obtained for them. In this case, because the magnetic field map is known, the evaluation of the predicted measurement function $h_i(\xk^i)$ from Equation \ref{eq:residual}, is equivalent to querying the magnetic anomaly map at the current estimated particle position $\xk^i$. The estimated position of the robot, $\xhat_m$, is approximated by the weighted average of all the particles $\xk^i$ in the sample $\mathcal{X}_k^s$.

The successive propagation of state samples, weight updating according to the computed residuals, and re-sampling, allows the filter to establish state-space regions where the vehicle is more likely to be located, and the distribution of the weights provides an approximation of the actual probability distribution of the state. Note that in the discussed scenario, the use of a particle filter for localization is convenient because 1) it admits multi-modal distributions that can arise due to the magnetic field isolines (same measurement for distinct $(x,y)$ positions), and 2) it allows for the use of a simple measurement model, in this case a table look-up of the magnetic field map. For the implementation of the particle filter we follow the particle filter mechanization described in \cite{thrun2002probabilistic}.

\section{Information-based Guidance}
This section goes into the details of the two information driven path planning algorithms. The observability-based path planning aims to increase the observability of the system state thereby driving the robot to regions with higher information content which in turn reduces estimation uncertainty. The second approach uses information entropy to measure the localization uncertainty and then chooses a control action so that the system uncertainty decreases.

\subsection{Non-Linear Observability-Based Guidance}\label{sec:ObservabilityGuidance}
The observability-based path planner generates paths using a receding-horizon-like dynamic programming (DP) approach which minimizes the euclidean distance to a given goal position $(x_{goal},y_{goal})$, while maximizing the local observability of the system. In particular, the DP path planning algorithm, inspired in \cite{yu2011observability}, uses the determinant of the observability Grammian $\mathcal{O}_G$ to quantify the degree of observability of the system for the next $p$-steps of the chosen horizon. Every time that the next $p$-steps ahead are planned, the optimal control to move the vehicle by only one step is applied. While the vehicle is in motion, a new optimal $p$-steps horizon is computed using the current position estimate, and the new optimal control command to move the robot to the next step is then executed to restart the process. The position estimate for this technique comes from the particle filter described in Sec~\ref{sec:particle_filter}.
The observability Grammian $\mathcal{O}_{G}$ is constructed as
$\mathcal{O}_{G}=\mathcal{O}_{nl}\trans\mathcal{O}_{nl}$, where $\mathcal{O}_{nl}$ is the nonlinear observability matrix. $\mathcal{O}_{nl}$ is obtained via the Lie derivatives, $L_{f}h$, of the system \eqref{eq:dynamics} and measurement model $h_i(\xhat_m)$. The computation of $\mathcal{O}_{nl}$ assumes that the magnetic measurement is equally affected by the vehicle in all directions, and thus, it is assembled with the gradients of the zeroth through $n-1$-th order Lie derivatives with respect to the position states only \cite{ramos2021observability}. This is,
\begin{equation}
\label{nonlinearObs}
\begin{array}{rcl}
    O_{nl} & = & \begin{bmatrix}
         dL_{\f^0}(h_1) \\
         dL_{\f^{1}}(h_1) \\
         \vdots \\
         dL_{\f^{n-1}}(h_1)
        \end{bmatrix}
\end{array}
\end{equation}
where the Lie derivatives are computed via 
\begin{equation}
L_{f}h(\boldsymbol{x})=\nabla h(\boldsymbol{x})\cdot \stdvec{f} =\sum\frac{\partial h}{\partial x_{i}}f_{i}    
\end{equation}
and $dL_{f}h(\boldsymbol{x})$ denotes the gradient of $L_{f}h(\boldsymbol{x})$.



For an observable system, $\mathrm{det}(\mathcal{O}_{nl}\trans\mathcal{O}_{nl}) \neq 0$. As the determinant of the Grammian matrix $\mathcal{O}_{G} \geq 0$, the path planning algorithm seeks to minimize $1/\mathrm{det}(\mathcal{O}_{G})$ to maximize observability. For the unicycle model \eqref{eq:dynamics}, the zeroth and first order Lie derivatives, $L_{f}^{0}h(\boldsymbol{x})$ and $L_{f}^{1}h(\boldsymbol{x})$, are given by
\begin{equation}
\mathcal{L} =\left[\begin{array}{c}
L_{f}^{0}\\
L_{f}^{1}
\end{array}\right] =\left[\begin{array}{c}
h(\x)\\
\frac{\partial h}{\partial x_{1}}Vcos\theta+\frac{\partial h}{\partial x_{2}}Vsin\theta
\end{array}\right]
\end{equation}
The observability matrix $\mathcal{O}_{nl}$ is then assembled as
\begin{align}
\mathcal{O}_{nl} =\frac{d\mathcal{L}}{dx_{i}} 
&=\left[\begin{array}{cc}
\frac{\partial h}{\partial x_1} & \frac{\partial h}{\partial x_2}\\
\left[\frac{\partial^2 h}{\partial x_1^2} \frac{\partial^2h}{\partial x_1 \partial x_2}\right] \cdot \stdvec{f} & \left[\frac{\partial^2 h}{\partial x_1 \partial x_2} \frac{\partial^2h}{\partial x_2^2}\right] \cdot \stdvec{f}
\end{array}\right]
\end{align}

Bringing all of the above components together, the path-planning algorithm plans $p$-steps ahead with initial condition $\x_{m_k}$ and tries to minimize a cost function J, by choosing the sequence input $\boldsymbol{\omega}= \{\omega_{i,} i=1,\dots, p\}$ corresponding to the $p$-steps horizon where $\omega_i \in \Omega$. In order to facilitate the DP approach the continuous action space $\Omega =\left[V,\omega\right]$ is replaced with a constant translational velocity and discrete rotational velocity. Thus, $\Omega$ becomes a set of discrete heading rates that can be commanded by the vehicle's controller. The optimal path can then be established by finding the optimal cost $J^*$ given by
\begin{equation}
     J^* \stackrel{\text{def}}{=} \min_{\boldsymbol{\omega}}\sum_{j=k+1}^{k+p-1}J(\stdvec{f}(\x_j,\omega_j))
\end{equation}
where,
\begin{equation}
    J=W_{goal}[(x_1-x_{goal})^2+(x_2-y_{goal})^2] + W_{obs}(\frac{1}{\text{det}\mathcal{O}_{G}})
\end{equation}
and $W_{goal}$ and $W_{obs}$ are user-selected weights to penilize the distance-to-goal and the observability cost.
By standard dynamic programming, the cost-to-go from time $k$, $J^*(\xk)$, satisfies the following recursion at any step in the horizon 
\begin{equation}
    J^*(\x_{k})=
    \min_{\boldsymbol{\omega_k}} (J(\stdvec{f}(\xk,\omega_k) + J^*_{k+1}(\x_{k+1}))
\end{equation}
where $J(\stdvec{f}(\xk,\omega_k)) + J^*_{k+1}(\x_{k+1})$ is the cost-to-go from $k+1$ and on.

\ifx false
\[
L_{f}^{1}h(\boldsymbol{x})=\nabla h(\boldsymbol{x}).f =\sum\frac{\partial h}{\partial x_{i}}f_{i}
\]

\[
\nabla h(\boldsymbol{x})=\left[\begin{array}{cc}
\frac{\partial h}{\partial x_{1}} & \frac{\partial h}{\partial x_{2}}\end{array}\right]
\]

\[
\frac{\partial h}{\partial x_{1}}=\frac{h(x_1+\Delta x_1,x_2)-h(x_1-\Delta x_1,x_2) }{2\Delta x_1}
\]

\[
L_{f}^{1}h(\boldsymbol{x})=\left[\begin{array}{cc}
\frac{\partial h}{\partial x_{1}} & \frac{\partial h}{\partial x_{2}}\end{array}\right]\left[\begin{array}{c}
vcos\theta\\
vsin\theta
\end{array}\right]
\]

\[
\mathcal{L} =\left[\begin{array}{c}
L_{f}^{0}\\
L_{f}^{1}
\end{array}\right] =\left[\begin{array}{c}
h(x_1,y_2)\\
\frac{\partial h}{\partial x_{1}}vcos\theta+\frac{\partial h}{\partial x_{2}}vsin\theta
\end{array}\right]
\]

\begin{align*}
O=\frac{d\mathcal{L}}{dx_{i}}=\left[\begin{array}{cc}
h_x & \frac{h(x,y+\Delta y)-h(x,y-\Delta y)}{2\Delta y}\\
\frac{h(x+\Delta x,y)-2h(x,y)+h(x-\Delta x,y)}{\Delta x^{2}}vcos\theta & \frac{h(x,y+\Delta y)-2h(x,y)+h(x,y-\Delta y)}{\Delta y^{2}}vsin\theta
\end{array}\right]
\end{align*}
\fi



\subsection{Expected Entropy Reduction Based Guidance}
In this paper, an entropy-based information-driven path planning algorithm is applied to the guidance problem by computing robot action to minimize localization uncertainty. This approach is developed based on \cite{burgard1997active} and shares the same assumption that the map of environment, which corresponds to the magnetic field, is known a priori. The approach considers the trade-off between minimizing the travel distance and improving the robot's localization uncertainty. Entropy is used to measure the localization uncertainty, and thus, expected entropy reduction (EER) is used to compute the best next action in path planning. The algorithm is developed as a one-step lookahead policy that can be applied in an online operation setting.

The probability distribution of the robot pose $\mathbf{x}_{k}$, or belief distribution, is denoted by $bel(\mathbf{x}_{k})$. The entropy of the distribution $bel(\mathbf{x}_{k})$ is defined by
\begin{equation}
    H(bel(\mathbf{x}_{k})) = -\int_{x} bel(\mathbf{x}_{k})\cdot\log (bel(\mathbf{x}_{k})) dx
\end{equation}
At each time step, the robot obtains a scalar magnetometer sensor measurement, $z_{k}$. Then, the belief distribution is updated by
\begin{equation}
    \label{eq:eer-sensormodel}
    bel(\mathbf{x}_{k}|z_{k})=\frac{f_{z_{k}}(z_{k}|x_{k})bel(\mathbf{x}_{k})}{p(z_{k})}
\end{equation}
The conditional probability $f_{z_{k}}(z_{k}|\mathbf{x}_{k})$ represents a sensor model. In this paper, the magnetometer is assumed to have a Gaussian noise with a variance, $\sigma$. Therefore, the distribution $f_{z_{k}}(z_{k}|x_{k})$ can be computed based on $\sigma$ and the map of environment, which is the given magnetic field.

Once the belief distribution is updated by $z_{k}$, the proposed algorithm computes the best next action based on the EER. In this paper, the control input is defined by an angular velocity. The robot dynamics follow \eqref{eq:dynamics} with Gaussian noise. Therefore, for a control input $u_{k}$, the belief distribution is updated by
\begin{equation}
    \label{eq:eer-motionmodel}
    bel(\mathbf{x}_{k+1}|\mathbf{x}_{k},u_{k}) = -\int f_{\mathbf{x}_{k+1}}(\mathbf{x}_{k+1}|\mathbf{x}_{k}, u_{k}) \cdot bel(\mathbf{x}_{k}) d\mathbf{x}_{k}
\end{equation}
where the distribution $f_{\mathbf{x}_{k+1}}(\mathbf{x}_{k+1}|\mathbf{x}_{k}, u_{k})$ represents the motion model. Therefore the EER value is computed by
\begin{multline}
    EER(bel(\mathbf{x}_{k}|z_{k}), u_{k}) = E_{z_{k+1}}[H(bel(\mathbf{x}_{k+1}|\mathbf{x}_{k}, u_{k}, z_{k+1}))] \\
    - H(bel(\mathbf{x}_{k}|z_{k}))
\end{multline}
where the expected entropy is computed based on eqs. \eqref{eq:eer-sensormodel} and \eqref{eq:eer-motionmodel}.

The EER-based algorithm considers the trade-off between the distance minimization and localization uncertainty reduction. Denoting the final goal location by $\mathbf{x}_{f}\in\mathbb{R}^{2}$, a function $Dist(\mathbf{x}_{f}, bel(\mathbf{x}_{k}|z_{k}), u_{k})$ is defined by the distance to the final goal position from the most probable current position, which is computed based on $bel(\mathbf{x}_{k+1}|\mathbf{x}_{k},u_{k})$. Therefore, by denoting the two user-chosen weights as $W_{obs}$ and $W_{goal}$, which correspond to the same weights defined in Section \ref{sec:ObservabilityGuidance}, the cost function is designed by
\begin{multline}
    J(bel(\mathbf{x}_{k}|z_{k}), u_{k}) = W_{obs} \cdot \big(\frac{1}{2}\big)^{EER(bel(\mathbf{x}_{k}|z_{k}), u_{k})} +\\ W_{goal}Dist(\mathbf{x}_{f}, bel(\mathbf{x}_{k}|z_{k}), u_{k})
    \label{eq:info_cost}
\end{multline}
in order to incorporate the scale of EER. By defining a finite and discrete action set, the action is chosen by
\begin{equation}
    u_{k} = \arg\min_{u_{k}}J(bel(\mathbf{x}_{k}|z_{k}), u_{k})
\end{equation}

\section{Results}\label{sec:results}
\begin{figure}[htbp]
          \begin{subfigure}{0.45\textwidth}\centering \includegraphics[width=0.9\textwidth]{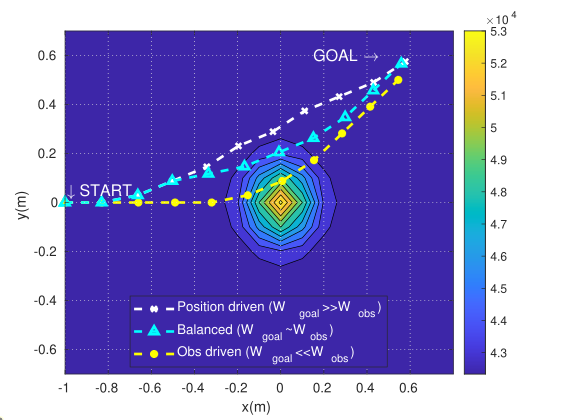}
          \caption{} 
          \label{f:1gaussian_world}
          \end{subfigure}
          \begin{subfigure}{0.45\textwidth}\centering \includegraphics[width=0.9\textwidth]{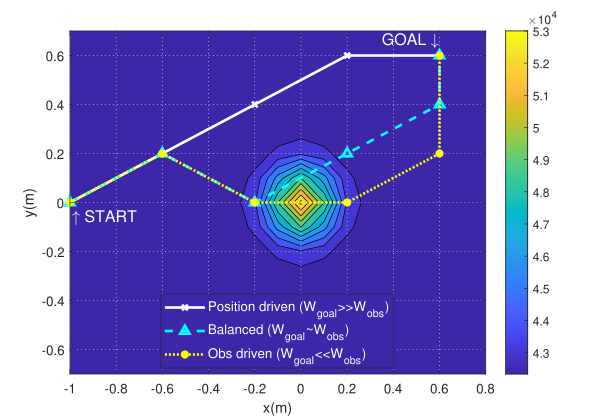}
          \caption{ } 
          \label{f:1gaussian_eer}
          \end{subfigure}
\caption{\label{f:virutal_sim}Paths from the virtual scenario when the ratio of $\frac{W_{obs}}{W_{goal}}$ is varied. Fig. (a) are paths from the non-linear observability-based path planning whereas (b) is from the expected entropy reduction technique. }
\end{figure}

Prior to using the real magnetic map data, both the observability- and information-based approaches were tested in a virtually ideal scenario. The intent was to validate path-planners in a controlled environment where resulting paths could be more predictable and insights on the approaches could be verified. The virtual scenario contains only one source of magnetic field information that is symmetric and placed at the origin of the virtual environment. The virtual scenario and some validation runs are shown in Figure \ref{f:virutal_sim}. In this figure it can be seen that both approaches showcase the same overall behavior: heavily penalizing the distance-to-go cost via $W_{goal}$, results in a fairly straight path as the optimization is mainly driven by the remaining distance to the goal position, as expected. On the other hand, penalizing the observability metric with a large $W_{obs}$ generates a path that attempts to leverage the rich information content from the magnetic field gradients, which was also expected. Finally, a more balanced penalization of the observability and distance-to-go terms ($W_{goal}\approx W_{obs}$), in turn, produced a path that still exploits the magnetic gradients for information, but less aggressively. Note that when the ratio $\frac{W_{obs}}{W_{goal}}$ becomes large, the final position of the vehicle can start to deviate from the goal position, as seen in Figure \ref{f:1gaussian_world}. It is also worth mentioning, that because the magnetic source is symmetric, and no other interacting fields are present, the paths are very well defined and the effect of the weights is clear. However, for more general magnetic sources shapes, the resulting paths are less discernible from one to the other, yet information gain was still observed as will be seen in the following section where a real magnetic map is used.
Both approaches used a simulated sensor with zero-mean Gaussian noise with standard deviation $\sigma=100$ nT. This noise value is based on the noise observed in the real sensor. The vehicle motion model is also perturbed with zero-mean Gaussian process noise to account for system mismodelling, imperfect actuators, and odometry errors. Specifically, a $1\sigma = 1$ cm error (per 5 cm) step was used for $x,y$ positions, and a $1\sigma=1$ deg error per (10 deg) step for $\theta$. Recall that it is considered that the vehicle moves with constant velocity $V$, and the steering angle $\omega$ is the input to the system.

\subsection{Simulations results}
\subsubsection{Observability-based approach}
\label{sssub:obs_sim}
\begin{figure}[thpb]
  \centering
  \includegraphics[width=0.9\columnwidth]{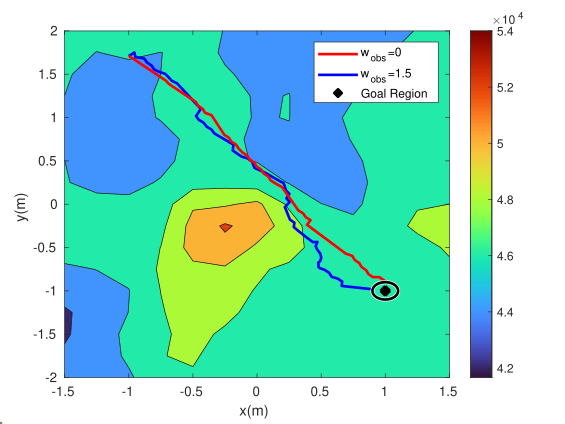}
  \caption{Paths of the robot navigating the magnetic anomaly map when using non-linear observability metric to plan paths shown in blue ($W_{obs}=1.5$), and when the metric is not used shown in red ($W_{obs}=0$).}
  \label{f:sim_mag_map_obs}
\end{figure}

\begin{figure}[thpb]
  \centering
  \includegraphics[width=0.9\columnwidth]{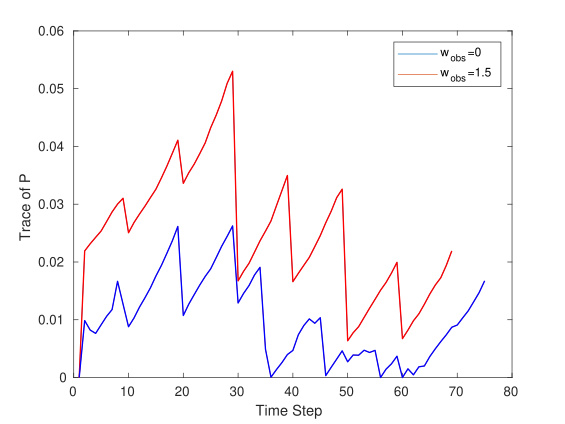}
  \caption{The trace of the position estimation uncertainty from the particle filter when the nonlinear observability metric is used to plan paths ($W_{obs}=1.5$), and when the metric is not used ($W_{obs}=0$).}
  \label{f:trace_obs_sim}
\end{figure}

The standard dynamic programming from the observabilty approach, uses a discretized control of $\omega \in[-45,-22,0,22,45]$ deg, a look-ahead horizon of 5 steps, and cost function weights of $W_{goal}=1$ and $W_{obs}=1.5$. The particle filter used 1000 particles and runs with the measurement and process noise parameters described above. The initialization of the particle filter uses samples from a Gaussian distribution with covariance $\stdvec{P}_0 = \text{diag}[0.1^2, 0.1^2, \text{deg2rad}(2)^2]$, with a mean at the simulated position of the robot. Whereas the simulation uses the real magnetic field map, the magnetic measurements are simulated by interpolating the map data at the current position and adding zero-mean Gaussian noise of $1\sigma=100$ nT.
Figure \ref{f:sim_mag_map_obs} shows the results for two simulated runs with the vehicle initial position at the top-left, and end-goal at the bottom-right. The most straight path corresponds to the case where the observability term is not considered for the path optimization ($W_{obs}=0$). This would correspond to a shortest distance approach that, although is using the magnetic map information for localization, does not purposefully leverage the magnetic gradients to minimize localization uncertainties. The other curve with larger deviations corresponds to the path generated with the observability term active ($W_{obs}=1.5$ and $W_{goal}=1$). In this second case, it can be seen that the path planner, as in the virtual scenario from Figure \ref{f:1gaussian_world}, is ``attracted" by the nearby magnetic field gradients while still reaching the goal region (circle with the dotted center).
Even when the paths are very similar in this simulation, it is seen that the system can largely benefit from gradient changes along the planned path. Figure \ref{f:trace_obs_sim} displays the system uncertainty via the trace of the particles sample covariance, where it can be seen that the overall uncertainty is smaller when the observability metric is optimized, and significant information gain can be achieved. Note, for example, that at approximately step 38 (around mid-way) when the vehicle travels in between two regions with locally higher gradients, the uncertainty collapses significantly with respect to a shortest-path approach.
Although the planned paths with and without the observability metric are similar, it is important to highlight that this also implies that the extra control effort necessary to gain information is not considerably different from that of the shortest-path approach. Furthermore, this may suggest that hybrid approaches (that can turn on and off the observability metric when ``worth it'') can also be beneficial in terms of control effort and computational savings.
\begin{figure}[thpb]
  \centering
  \includegraphics[width=0.9\columnwidth]{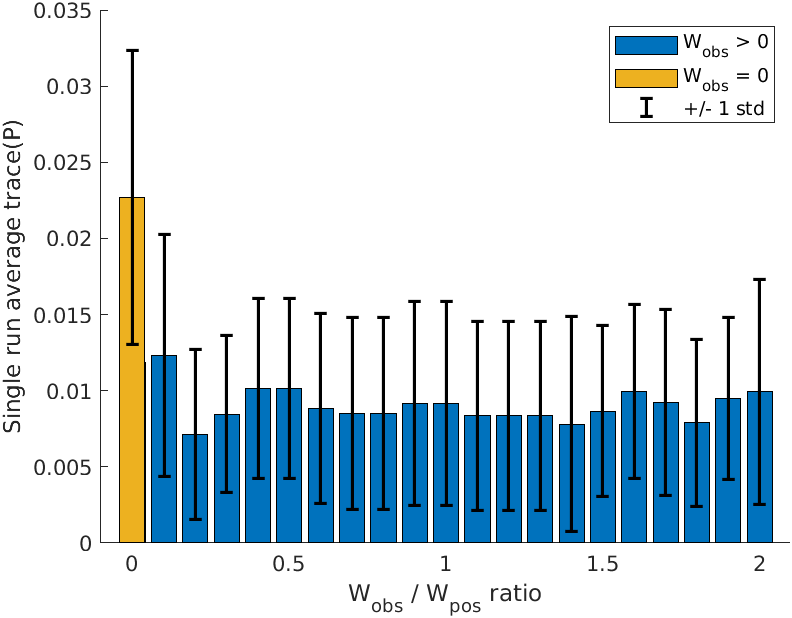}
  \caption{Single-run average over time of trace of the estimation uncertainty ($\pm$ 1 standard deviation) for different $W_{obs}/W_{goal}$ ratios.}
  \label{f:bars_obs_ratios}
\end{figure}
Figure \ref{f:bars_obs_ratios} shows the mean trace of the sample covariance when the planning algorithm uses different $W_{obs}/W_{goal}$ ratios. In this figure, the $\pm 1$ standard deviation is also shown. The ratios beyond $W_{obs}/W_{goal} > 2$ are not plotted but they were seen to be less effective as the algorithm starts to heavily prioritize navigation across the magnetic field gradients instead of arriving at the goal point. Overall, Figure \ref{f:bars_obs_ratios} shows the benefit of leveraging the observability information into the path planning, which is directly reflected on the reduction of the state uncertainty (while still arriving to the goal pose). Although different $W_{obs}/W_{goal}$ ratios seemed to have approximately the same uncertainty reduction, the effect of the weights ratio was seen to be dependent on the region of the magnetic field being navigated. Nonetheless, the observability-aware method was always seen to be have less uncertainty estimation than a straight-path approach.

\subsubsection{EER-based approach}
The EER-based planner is implemented in a one-step look-ahead fashion, where the planner computes the best angular velocity from $\omega\in[-40, -20, 0, 20, 40]$ (deg/s) based on the cost function at every time step and then the robot translates with a constant velocity at every time step. The EER-based approach uses a grid of 0.25 m by 0.25 m in $x$ and $y$ axes. Note that due to the scale of EER values, the effects from different $W_{obs}$ and $W_{goal}$ ratios may not be identical to the observability-based method. The parameters $\stdvec{P}_0$ and $1\sigma$ are set identical to the ones in \ref{sssub:obs_sim}.

\begin{figure}[thpb]
  \centering
  \includegraphics[width=0.9\columnwidth]{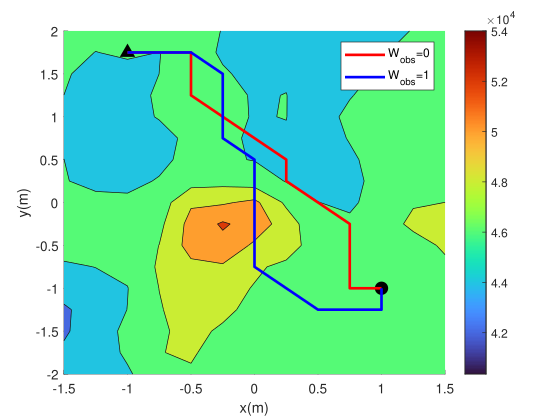}
  \caption{Paths of the robot navigating the magnetic anomaly map when using expected entropy reduction metric.}
  \label{f:sim_mag_map_eer}
\end{figure}

\begin{figure}[thpb]
  \centering
  \includegraphics[width=0.9\columnwidth]{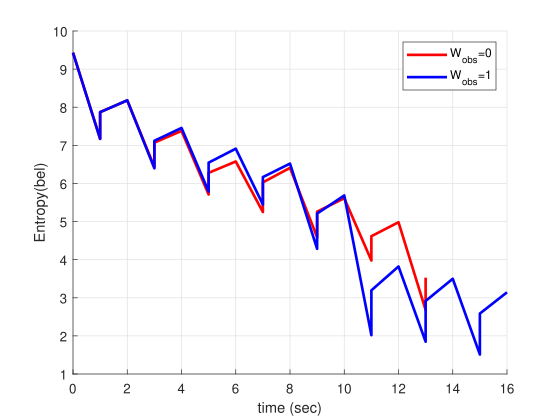}
  \caption{The entropy of the position estimation from the belief function over time corresponding to the paths of Fig. \ref{f:sim_mag_map_eer}.}
  \label{f:info_entropy_time}
\end{figure}

\begin{figure}[thpb]
  \centering
  \includegraphics[width=0.9\columnwidth]{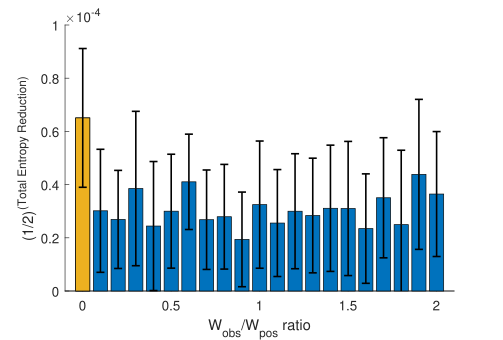}
  \caption{Single-run average of total entropy reduction ($\pm$ 1 standard deviation) for different $W_{obs}/W_{goal}$ ratios.}
  \label{f:info_bar}
\end{figure}

Figure \ref{f:sim_mag_map_eer} shows that the path planned considering EER (blue line) takes a longer path in order to obtain more information on localization, while the path that does not considers the EER (red line) took the shortest path towards the goal position. The entropy of the belief function at each time step is shown and compared in Fig. \ref{f:info_entropy_time}. The entropy drops are due to the observations, while the increments are due the motion model uncertainty. In this simulation, after the EER-based planner reduces the entropy of the belief function enough, the EER values of the control inputs do not differ as at the beginning because the belief state has sufficiently converged. With only small reductions in the remaining entropy, the distance to the goal position gives a higher impact in the cost function, driving the robot to the goal position.
In Fig. \ref{f:info_bar}, the entropy reduction from Eq. \eqref{eq:info_cost} is shown for a range of $W_{obs}/W_{pos}$ ratios. For each ratio, 20 sample paths were generated and the mean and $\pm 1\sigma$ is plotted. Fig. \ref{f:info_bar} shows that the entropy reduction is higher when the planner considers the entropy cost. The results does not show a significant correlation between the ratio and the amount of total entropy reduction; however, this relationship can be further investigated with different features in magnetic field, higher resolution of the magnetic field map, and longer horizon path planning.
\subsection{Hardware}
In order to validate the algorithm described in Sec.~\ref{sec:ObservabilityGuidance}, the nonlinear observability-based guidance law is deployed on a Turtlebot 3 (shown in Fig.~\ref{fig:robotic_plaform}). The ground robot runs the particle filter (with 1000 particles) on a on-board Intel i5 NUC and uses an optically pumped TwinLeaf MicroSAM scalar magnetometer to obtain measurements at 0.25 Hz. The particle filter and guidance law are implemented in C++ and used the Robot Operating System (ROS) to enable message passing between the varies sub-systems. 

The robot travels at a constant velocity of 0.2 m/s, the optimal control computes the yaw rate ($\omega$) of the robot and can choose between three discrete options -10 deg/sec, 0, 10 deg/sec. A motion capture system serves as a stand-in to initialize the filter and to provide delta-pose measures to propagate the filter. The particle filter processes magnetic measurements at approximately 0.25 Hz, but this rate is increased based on the observability metric, i.e., when the vehicle is in a zone with high information content, the filter reads the magnetometer at 0.5 Hz.

\begin{figure}[thpb]
  \centering
  \includegraphics[width=0.9\columnwidth]{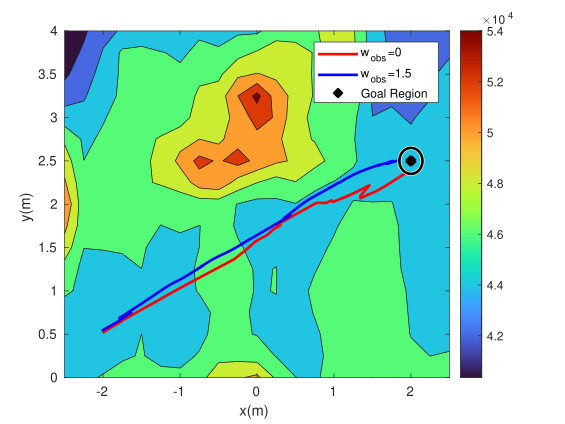}
  \caption{Pah followed by the ground robot overlayed on the magnetic map. The two paths show when the observability-based path planning is used ($W_{obs}$ = 1.5) and when it is not ($W_{obs}$ = 0).}
  \label{f:hardware_mag_map}
\end{figure}

\begin{figure}[thpb]
  \centering
  \includegraphics[width=0.9\columnwidth]{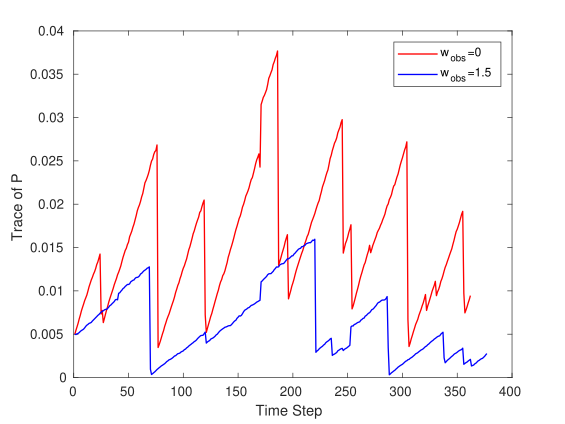}
  \caption{The trace of the localization uncertainty produced by the particle filter when the observability-based path planning is utilized ($W_{obs}$ = 1.5), when its turn off ($W_{obs}$ = 0). Generated from the path followed in Fig.~\ref{f:hardware_mag_map} }
  \label{f:tracep}
\end{figure}

Fig.~\ref{f:hardware_mag_map} shows the path followed by the ground robot over a run where it is commanded to (2m, 2.5m) from a start position near (-2m, 0.5m). To show the benefits of using an information-based path planning technique, each path is run twice: with and without the observability-based guidance. When $W_{obs}$ is set to 1.5, we notice that the robot is pulled closer to the region with more magnetic features starting at around ($x$ = 0.25m, $y$ = 1.75m), compared to when $W_{obs}$ is set to zero. The paths contain small jumps that result from the particle filter correcting the estimated position of the robot after receiving a measurement. The primary causes for the estimation error arises from two main sources: 1) the scalar magnetometer is susceptible to noise which adversely affects the particle filter especially in regions where the magnetic gradients do not significantly change and 2) the heading compensation used when generating the map does not capture all of the heading-dependent effects present in the magnetic field.
Fig. ~\ref{f:tracep} shows the trace of the sample covariance produced by the particle filter for the two paths shown in Fig.~\ref{f:hardware_mag_map}. It is also noted that the average estimation error for the run when $W_{obs}=0$ is 32cm vs 23cm when $W_{obs}=0$. From both Fig. ~\ref{f:tracep} and the estimation error, it is evident that the observability-based path planning helps increase navigational accuracy by steering the robot to regions with more magnetic anomaly features.
\section{Conclusion}
We have presented two information aware guidance algorithms for autonomous agents using magnetic anomalies to navigate in a GPS-denied environment. One based on nonlinear observability and the second based on the expected entropy reduction; through simulations that use the magnetic map of our laboratory space, we show that the two techniques improve navigational accuracy. Additionally, hardware experiments of the observability-based technique are deployed on a ground robot to demonstrate the benefits of this information aware guidance law. While this article specifically focuses on magnetic anomaly based navigation, the same techniques can be applied to other map-based navigation systems such terrain navigation. 


\section{Acknowledgments}
The authors would like to thank Jared Paquet for supporting the hardware experimentation and Kristy Waters for providing valuable feedback on this manuscript. The assistance of Vaishnav Tadipadpthi, Phillip Mitchell, and Anubhav Gupta, was invaluable in developing the particle filter-based localization and in collecting the magnetic anomaly map.




\bibliographystyle{IEEEtran}
\bibliography{MagNav}
\end{document}